\documentclass[pre, preprint, groupedaddress, showkeys,  showpacs, amsmath]{revtex4-1}
\usepackage[dvipdfm]{graphicx}

\begin{document}

%\preprint{}

\title{Nonadiabatic Quantum Annealing for One-Dimensional Trasverse-Field Ising Model}


\author{Hitoshi Katsuda}
\email[]{katsuda.phys.edu@gmail.com}
\author{Hidetoshi Nishimori}

%\thanks{}
\affiliation{Department of Physics, Tokyo Institute of Technology,
Oh-okayama, Meguro-ku, Tokyo 152-8551, Japan}

\date{\today}

\begin{abstract}
We propose a nonadiabatic approach to quantum annealing, in which we repeat quantum annealing in nonadiabatic time scales, and collect the final states of many realizations to find the ground state among them.
In this way, we replace the diffculty of long annealing time in adiabatic quantum annealing by another problem of the number of nonsidabatic (short-time) trials.
The one-dimensional transverse-field Ising model is used to test this idea, and it is shown that nonadiabatic quantum annealing has the same computational complexity to find the ground state as the conventional adiabatic annealing does. This result implies that the nonadiabatic method may be used to replace adiabatic annealing to avoid the effects of external disturbances, to which the adiabatic method is more prone than the nonadiabatic counterpart.
\end{abstract}


\maketitle
\newcommand{\up}{\uparrow}
\newcommand{\down}{\downarrow}
\newcommand{\J}{\{J_{ij}\}}
\newcommand{\etas}{\{\eta_i\}}
\newcommand{\Tr}{\text{Tr}}
\newcommand{\spin}{\mathbf{S}}
\newcommand{\Schr}{Schr\"{o}dinger }
\newcommand{\QT}{\mathcal{O}_{\tau}}
\newcommand{\xis}{\{\xi\}}
\newcommand{\RFJ}{e^{-\beta W}\rangle_{h_0\rightarrow h_{\tau}}}
\newcommand{\JQA}{\frac{t}{\tau}J}
\newcommand{\GQA}{\biggl(1-\frac{t}{\tau}\biggr)\Gamma}

% References should be done using the  \cite,  \ref, and \label commands
\section{Introduction}\label{sec: Introduction}
Quantum annealing (QA) is an algorithm to solve optimization problems by the method of quantum mechanics and statistical physics \cite{QA1, QA2, QA3, QA4, QA5, QA6, QA7, QA8}. A combinatorial optimization problem can often be expressed as a problem to find the ground state of a spin system, where the cost function is described as the Hamiltonian \cite{opt1, opt2, opt3}. Most current research activities on QA are based on the adiabatic theorem of quantum mechanics \cite{adiabatic}. 
The adiabatic theorem states that, if the time evolution of the Hamiltonian is sufficiently slow, the quantum system stays in the instantaneous ground state when we start from the ground state. Therefore, if we choose the initial Hamiltonian to have a trivial ground state and the final Hamiltonian as the target Hamiltonian whose ground state we are to find, the solution is automatically reached after the adiabatic time evolution following the time-dependent Schr\"odinger equation.

To describe the problem more quantitatively, let us consider the following time-dependent Hamiltonian
\begin{equation}
H(t)=\biggl( 1-\frac{t}{\tau}\biggr) H_0 + \frac{t}{\tau}V.
\label{Hamiltoian}
\end{equation}
Initially ($t=0$), the Hamiltonian $H(0)$ equals to $V$ with a trivial ground state, and at the final time $t=\tau$ the Hamiltonian $H(\tau)$ reduces to the target Hamiltonian $H_0$. $\tau$ determines the time scale of a single realization of annealing. In order to apply the adiabatic theorem, the following inequality has to be satisfied:
\begin{equation}
1\gg \frac{\max_{0\leq t \leq \tau}[ \langle 1(t)| \frac{dH(t)}{dt}|0(t)\rangle]}{\min_{0\leq t \leq \tau}[\Delta (t)^2]},
\label{adiabatic_condition_general}
\end{equation}
where $|0(t)\rangle $ and $|1(t)\rangle $ express the instantaneous ground and first excited states of the Hamiltonian at time $t$, respectively, and $\Delta$ describes the energy gap between those states.

In practice, the minimum energy gap $\min_{0\leq t \leq \tau}[\Delta (t)]$ often increases rapidly as the system size $N$ increases, and correspondingly, the annealing time $\tau$ diverges. Especially, for problems called NP hard  \cite{NP} , the minimum energy gap decreases exponentially with $N$. This difficulty of adiabatic quantum annealing is our motivation to follow a nonadiabatic approach to quantum annealing.

In the present paper, we propose a nonadiabatic method to find the ground state of the target Hamiltonian. For this purpose, we repeat the realizations of quantum annealing, and a single realization is done in 
a nonadiabatic time scale. We collect the final states of many realizations with the expectation that the ground state is found among them. In this way, we replace the difficulty of long annealing time in adiabatic 
quantum annealing by another problem of the number of nonadidabatic (short-time) trials.

We show that nonadiabatic quatum annealing can succeed in the same time scale (computational complexity) as the adiabatic one for the one-dimensional transverse-field Ising model.  
Nonadiabatic QA has the advantage of an easier practical implementation because we do not have to tune the time-dependent parameters carefully to keep the system in the ground state. This fact also implies that the nonadiabatic method may be more robust against external disturbances.
In fact, in an experimental implementation of quantum annealing, Johnson et al  \cite{QAExp1} repeated annealing processes and measured the probability that the final system is in the ground state 
(See also  \cite{QAExp2}).

This paper is organized as follows. 
In the next section, we formulate the idea explicitly for the transverse-field Ising chain and analyze the system analytically and numerically. The final section is for summary and discussion.

In the whole of this paper, we use the units $k_B=\hbar=1$.

\section{Nonadiabatic Quantum Annealing}\label{Sec: Nonadiabatic Quantum Annealing}

Let us introduce the one-dimensional transverse-field Ising model:
\begin{equation}
H(t)=-\frac{t}{\tau}\sum_{i=1}^N \sigma_i^z\sigma_{i+1}^z -\biggl( 1-\frac{t}{\tau}\biggr)\sum_{i=1}^N\sigma_i^x.
\label{eq:one-dimensional transverse-field Ising model}
\end{equation}
The Pauli matrix $\sigma_i^{x, z} $ satisfies the periodic boundary condition $\sigma_{N+1}^{x, z} =\sigma_1^{x, z}$.
The first coupling term corresponds to the target Hamiltonian $H_0$ in Eq. (\ref{Hamiltoian}), and the second term is the initial Hamiltonian $V$.

Following the standard procedure  \cite{Dziarmaga1, Dziarmaga2}, the Hamiltonian is decomposed into the sum of the two-state space of each mode $q = \pm \pi /N, \pm 3\pi /N,  \dots, \pm (N-1)\pi /N$. The time-dependent Schr\"odinger equation is reduced to a simple equation
\begin{equation}
i\frac{d}{dt}
\biggl( \begin{matrix}
u_q\\v_q
\end{matrix} \biggr)
=2\begin{pmatrix}
(1-t/\tau ) -(t/\tau)   \cos q & (t/\tau)  \sin q \\ (t/\tau)  \sin q & -(1-t/\tau ) +(t/\tau)  \cos q 
\end{pmatrix}
\biggl( \begin{matrix}
u_q\\v_q
\end{matrix} \biggr).
\label{eq: reduced Bogoliubov-de Gennes equation}
\end{equation}
The adiabatic condition, applied to the smallest gap among various $q$ values, turns out to be
\begin{equation}
\tau \gg \frac{N^2}{2\pi^2}.
 \label{eq:reduced adiabatic condition}
\end{equation}
The energy gap $\Delta_q(t)$ takes its minimum at $t=\tau/2$ for any $q$.

According to Vitanov and Garraway  \cite{VG}, the dynamics of such a two-level quantum system is exactly solvable and the solution is expressed as a linear combination of parabolic cylinder functions $D_{\nu}(z)$. 
However, we follow an approximate method, which is expected to be asymptotically correct in the limit 
$N \gg 1$ and is easier to see explicitly.

\begin{figure}[h]
\includegraphics[scale=0.5]{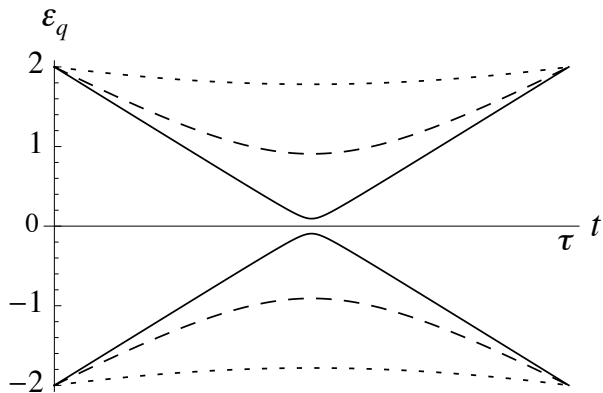}
\caption{Energy spectra $\pm \epsilon_q$ for $q=0.7\pi, 0.3\pi, 0.03\pi$ from the outer curves to the inner. Tunneling happens only around $t=\tau/2$ and only for small $q$.}
\label{fig: energy spectra.}
\end{figure}

First, we show the energy spectra $\pm\epsilon_q$ for $q=0.7\pi, 0.3\pi, 0.03\pi$ in Fig.  \ref{fig: energy spectra.}. It is observed that a state transition is likely to happen only around $t=\tau/2$, where the gap is very small for small $q$.
This fact allows us to replace $t/\tau $ in the off-diagonal elements in Eq. (\ref{eq: reduced Bogoliubov-de Gennes equation}) by $1/2$. Then the problem reduces to the Landau-Zener type  \cite{Landau, Zener}.

Next, following Dziarmaga  \cite{Dziarmaga1}, we shift the time $t\rightarrow t'$,
\begin{equation}
t'=\frac{\tau \sin q}{1+\cos q}\biggl(1-\frac{t}{\tau}(1+\cos q)\biggr).
\label{eq: time transformation}
\end{equation}
Equation (\ref{eq: reduced Bogoliubov-de Gennes equation})  is then transformed to
\begin{equation}
i\frac{d}{dt'}
\biggl( \begin{matrix}
u_q\\v_q
\end{matrix} \biggr)
=\begin{pmatrix}
-\mu_q t' & -1 \\ -1 & \mu_q t'
\end{pmatrix}
\biggl( \begin{matrix}
u_q\\v_q
\end{matrix} \biggr),
\label{eq: transformed Bogoliubov-de Gennes equation}
\end{equation}
where $\mu_q =2(1+\cos q)/(\tau\sin ^2 q)$. 
For a positive $q$, the transformed time $t'$ runs in the range
\begin{equation}
-\frac{2}{\sin q}\mu_q^{-1}\leq t' \leq \frac{2}{\tan q}\mu_q^{-1}.
\label{eq: range of t'}
\end{equation}
For negative $q$, the signs of inequalities are reversed. Tunneling takes place during
$- \mu_q^{-1}\leq t' \leq \mu_q^{-1}$ because the off-diagonal elements of Eq. (\ref{eq: transformed Bogoliubov-de Gennes equation}) are then larger than the diagonal elements in the absolute value.
This latter time window is  included in the above range of $t'$, Eq. (\ref{eq: range of t'}) for small $|q| \lesssim\pi /3$.
We are therefore justified to extend the range of $t'$ to $(-\infty, \infty)$ for small $q$ because there is no transition in the extra time range then included.
Then Eq. (\ref{eq: transformed Bogoliubov-de Gennes equation}) becomes equivalent to the Landau-Zener equation  \cite{Landau}. The Landau-Zener formula gives the transition probability $P_q$ for each $q$,
\begin{equation}
P_q =\exp\biggl[\frac{-2\pi (-1)^2}{2\mu_q} \biggr]= \exp\biggl[\frac{- \pi \tau\sin ^2 q}{2(1+\cos q)} \biggr].
\label{eq: Landau-Zener formula}
\end{equation}
For larger $q$, transitions do not happen.

Let us next estimate the probability to obtain the ground state $P_{\text{GS}}$ in nonadiabatic quantum annealing. 
Similarly to the adiabatic case, we set the initial state as the ground state of the initial Hamiltonian, $(u_q, v_q)=(0, 1)$. Then we have
\begin{equation}
\begin{split}
P_{\text{GS}} = \prod_{|q|\leq \pi/3}\biggl( 1-\exp\biggl[\frac{-\pi\tau\sin ^2 q}{2(1+\cos q)} \biggr] \biggr).
\label{PGS}
\end{split}
\end{equation}
It is convenient to take a logarithm,
\begin{equation}
\begin{split}
\ln P_{\text{GS}} & = \sum_{|q|\leq\pi /3}\ln \biggl( 1-\exp\biggl[\frac{-\pi\tau\sin ^2 q}{2(1+\cos q)} \biggr] \biggr)\\
&\rightarrow \frac{N}{2\pi}\int_{-\frac{\pi}{3}}^{\frac{\pi}{3}}dq\ln\biggl[ 1-\exp\biggl( \frac{-\pi\tau\sin ^2 q}{2(1+\cos q)}  \biggr)  \biggr]\\
 &=-\frac{N}{2\pi}\sum_{n=1}^{\infty}\frac{1}{n}  \int_{-\frac{\pi}{3}}^{\frac{\pi}{3}}dq\exp\biggl( \frac{-n\pi\tau\sin ^2 q}{2(1+\cos q)}  \biggr).\\
\label{log PGS}
\end{split}
\end{equation}
Since nonadiabatic transitions happen only for $q \approx 0$ having small energy gaps,  the following approximation is allowed,
\begin{equation}
\exp\biggl( \frac{-n\pi\tau\sin ^2 q}{2(1+\cos q)}  \biggr)\approx \exp\biggl(-\frac{n\pi \tau}{4}q^2 \biggr).
\label{eq: Taylor Pq (LZ)}
\end{equation}
Additionally, if we extend the integration range from $(-\pi/3, \pi/3)$ to $(-\infty, \infty)$ in Eq. (\ref{log PGS}),  the integration becomes Gaussian, which is valid for large $\tau$ since the integrand is very small outside the range $(-\pi /3, \pi /3)$. We then have a simple form of $\ln P_{GS}$:
\begin{equation}
\ln P_{GS}\approx -\frac{N}{\pi \sqrt{\tau}}\sum_{n=1}^{\infty}\frac{1}{n^{3/2}}
 = -\frac{N}{\pi \sqrt{\tau}}\zeta (3/2),
\label{eq: explicit form of log PGS}
\end{equation}
where, $\zeta (z)$ is the Riemann zeta function,
\begin{equation}
\zeta (z) =\sum_{n=1}^{\infty}\frac{1}{n^z},
\label{eq: Riemann zeta function}
\end{equation}
with $\zeta (3/2) \approx 2.61248$. 

As is clear from the above equation, $P_{\text{GS}}$ is a monotone decreasing function of the annealing time $\tau$, which seems to be physically trivial. However, we repeat nonadiabatic processes many times, and obtain the ground state. Thus, the time necessary to obtain the ground state is the product of the annealing time $\tau$ and the number of trials. On average, the number of trials converges to the inverse of the probability $P_{\text{GS}}$. For example, if $P_{GS}=1/3$, we have to repeat the process three times on average to reach the ground state.
Therefore, we have to consider the quantity $\tau/P_{\text{GS}}$ as the correct measure of computational complexity. We then again analyze,
\begin{equation}
\ln\frac{\tau}{P_{\text{GS}}}=\ln \tau +\frac{N}{\pi \sqrt{\tau}}\zeta(3/2).
\label{eq:log tau / PGS}
\end{equation}
If we differentiate the above equation with respect to the annealing time $\tau$, we find that the above function takes its minimum at finite $\tau$, 
\begin{equation}
\tau=\frac{N^2}{4\pi^2}\biggl(\zeta(3/2) \biggr) ^2.
\label{eq: tau gives minimum annealing time}
\end{equation}
We substitute this $\tau$ into the right hand side of Eq. (\ref{eq: explicit form of log PGS}) and obtain the minimum value of the time necessary to obtain the ground state,
\begin{equation}
\min \frac{\tau}{P_{\text{GS}}}= \biggl( \frac{N}{2\pi}\zeta (3/2) \biggr)^2 e^{2} \approx 1.27732 \times N^2.
\label{eq: minimum time to find the ground state}
\end{equation}

It is also possible to estimate the standard deviation of the realization times, in addition to the average $1/P_{GS}$, using Eqs. (\ref{log PGS}), (\ref{eq: tau gives minimum annealing time}) and (\ref{eq: minimum time to find the ground state})  as,
\begin{equation}
 \sigma = \sqrt{e^2(1+e^{-4\beta})^N(1+e^2(1+e^{-4\beta})^N)}.
\label{eq: fluctuation of the realization times}
\end{equation}
The time  to reach the ground state may then be the product of $\tau$ and $1/P_{\text{GS}} + \sigma$, which does not change the computational complexity $\mathcal{O}(N^2)$. Therefore we have shown that nonadiabatic quatum annealing can succeed in the same time scale as the adtiabatic one, Eq. (\ref{eq:reduced adiabatic condition}), for the one-dimensional transverse-field Ising model.

\begin{figure}[h]
\centering
\includegraphics[scale=0.45]{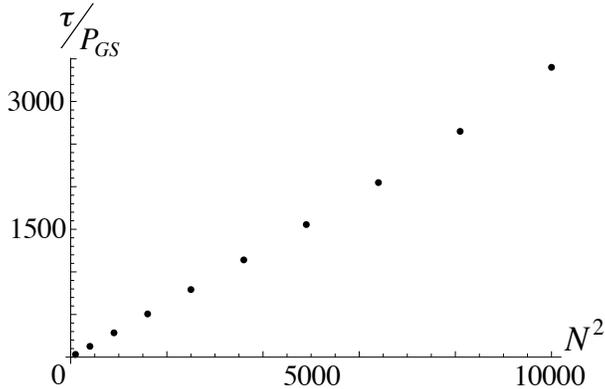}
\caption{Numerical results for $N^2$ versus $\tau/P_{GS}$ for $N=10, 20, \dots, 100$.}
\label{fig:timescale}
\end{figure}

In order to check the validity of a few approximations in the analyses around Eqs. (\ref{eq: Landau-Zener formula}) and (\ref{eq: Taylor Pq (LZ)}), we numerically solved the Scr\"odinger equations for each mode (\ref{eq: reduced Bogoliubov-de Gennes equation}) and calculated the time necessary for nonadiabatic quantum annealing $\tau /P_{\text{GS}}$ for several single realization time $\tau$ and the number of spins $N$. Figure  \ref{fig:timescale}. shows that the total annealing time $\tau/P_{GS}$ is propotional to $N^2$ for $N=10, 20, \dots, 100$. This observation supports the result Eq. (\ref{eq: minimum time to find the ground state}) that the computational complexity is $\mathcal{O}(N^2)$.

\section{Summary and Discussion}
In the present paper, we proposed the method of nonadiabatic quantum annealing. We repeat trials of quantum annealing in the nonadiabatic time scale, and find the ground state among them. For the one-dimensional transverse-field Ising model, we analytically estimated the computational complexity of nonadiabatic quantum annealing and have shown that the computational complexity is the same as the adiabatic one $\mathcal{O}(N^2)$. In order to the check the validity of our analyses, we numerically solved the Schr\"odinger equation, and found consistent results.

As mentioned in Sec.  \ref{sec: Introduction}, this is a first step to open the door of nonadiabatic quantum annealing. We have to study other models, especially models having randomness \cite{opt1, opt2, opt3}. It would also be useful to investigate the effects of thermal agitation in some details.

\begin{acknowledgments}
HK is grateful for financial supports provided through \textit{Research Fellowships of the Japan Society for the Promotion of Scientists}, \textit{The 21st Century Global COE Program at Tokyo Institute of Technology}, and \textit{CREST, JST}.
\end{acknowledgments}

\end{document}